\documentclass[11pt]{article}

\usepackage{times}
\usepackage{natbib}
\bibpunct{(}{)}{;}{a}{,}{,}

\usepackage{fullpage}
\usepackage{graphicx}
\usepackage{amsmath,amssymb,amsthm}
\usepackage{bm}
\usepackage{latexsym}
\usepackage{hyphenat}

\usepackage[pdftex,colorlinks=true,linkcolor=blue,citecolor=blue,urlcolor=blue,bookmarks=false,pdfpagemode=None]{hyperref}

\usepackage{url}
\makeatletter
\def\url@leostyle{%
  \@ifundefined{selectfont}{\def\UrlFont{\sf}}{\def\UrlFont{~\it\small\ttfamily}}}
\makeatother
\usepackage{verbatim}
\usepackage{fancyhdr}
\usepackage{setspace}
\usepackage{paralist}
\usepackage{boxedminipage}

\newcommand{\bv}{\begin{array}}
\newcommand{\ev}{\end{array}}
\newcommand{\bit}{\begin{itemize}}
\newcommand{\eit}{\end{itemize}}
\newcommand{\ben}{\begin{enumerate}}
\newcommand{\een}{\end{enumerate}}
\newcommand{\beq}{\begin{equation}}
\newcommand{\eeq}{\end{equation}}
\newcommand{\bvq}{\begin{eqnarray}}
\newcommand{\evq}{\end{eqnarray}}

\begin{document}
\thispagestyle{empty}
\section*{\centerline{\Large Getting started in probabilistic graphical models}\vspace{20pt}\newline 
          \it \normalsize 
          Edoardo M. Airoldi~\footnote{Address correspondence to: Edo Airoldi, Carl Icahn Laboratory, Princeton University, Princeton, NJ 08544, USA.}\hfill (eairoldi@princeton.edu)\newline
          Lewis-Sigler Institute for Integrative Genomics ~\&~
          Department of Computer Science \newline
          Princeton University}

\begin{abstract}
Probabilistic graphical models (PGMs) have become a popular tool for computational analysis of biological data in a variety of domains. But, what exactly are they and how do they work? How can we use PGMs to discover patterns that are biologically relevant? And to what extent can PGMs help us formulate new hypotheses that are testable at the bench? This note sketches out some answers and illustrates the main ideas behind the statistical approach to biological pattern discovery.
\end{abstract}



\paragraph{Overview.}

Probabilistic graphical models offer a common conceptual architecture where biological and mathematical objects can be expressed with a common, intuitive formalism. This enables effective communication between scientists across the mathematical divide by fostering substantive debate in the context of a scientific problem, and ultimately facilitates the joint development of statistical and computational tools for quantitative data analysis.
 A number of success stories have appeared over the years \citep{Fels:1981,Prit:Step:Donn:2000,Frie:2004,Xing:Karp:2004}. Today, probabilistic graphical models promise to play a major role in the resolution of many intriguing conundrums in the biological sciences.
 The goal of this short article is to be a dense, informative introduction to {\em the language} of probabilistic graphical models, for beginners, with {\em pointers} to successful applications in selected areas of biology. The exposition introduces the essential concepts involved in PGMs in the context of the various stages of a typical collaboration between natural and computational scientists, and discusses the aspects to which each scientist should contribute to carry out the data analysis successfully using PGMs.

Let us start by considering a specific problem in transcriptional regulation. Given measurements about the abundance of gene transcripts in retinal cells across stages of development, we would like to discover which functional processes are relevant for development, and reveal which ones are most important at which stage.
 To develop a PGM to address this problem, we begin by identifying the biological objects that would appear in a cartoon model of how cellular development impacts transcription. In this illustrative example, we have genes and functional processes/contexts. It is reasonable to assume that each gene will participate in multiple functional processes, although typically in a small number of them, and that not all functional processes will be important at all stages of development. We then assess what aspects of the problem we can probe directly, with experimental techniques, and what aspects we cannot. In the example, while an abundance of gene transcripts can be obtained, for instance, via SAGE (serial analysis of gene expression), it is harder to measure functional processes. However, the latter could be operationally defined as sets of genes that share a similar temporal regulation pattern; this definition has the advantage of creating a connection between membership of genes to functional processes (i.e., an unobservable mapping) and similarity of the temporal expression profiles (i.e., observable quantities). The establishment of connections between those biological objects that we can probe and those that we cannot ends a first conceptual effort.
 
 A cartoon model of how cellular development impacts transcription is now specified in terms of genes and their abundance, functional processes, and membership of genes to functional processes. Next we translate the biological players and the connections we established among them into mathematical quantities (i.e., random variables) and connections among them (i.e., statistical dependencies). This translation specifies the model structure. At this stage, we rely on biological intuitions to fine-tune the model, for instance, by deciding which sources of variability in the measurements carry information about the latent variables and which do notÑif the temporal expression profiles of genes A and B are similar on a relative scale, but their absolute abundance is quite different, should we believe that they both participate in the same functional processes? Last, we assign numerical values to those quantities that are unknown in the final model specifications (i.e., we fit the model to the data) and we use them to develop biological intuitions in the context of the original problem. (Functional aspects of retinal development, in mouse, are fully addressed in \citealp{Airo:Fien:Xing:2006}.)
  
In the following, we briefly introduce the basic mathematical quantities that enable the translation of a cartoon model of biology into a PGM, and we review strategies to assign numerical values to the unknown quantities underlying any PGM that are most likely given the observations. We conclude with an overview of selected applications, complete with pointers to published work.

\paragraph{The Basics.}

A probabilistic graphical model defines a family of probability distributions that can be represented in terms of a graph. Nodes in the graph correspond to random variables; its structure translates into statistical dependencies (among such variables) that drive the computation of joint, conditional, and marginal probabilities of interest \citep{Jord:2004}. In applications, most of the (node-specific) random variables are chosen to express the variability of an observed quantity, such as the expression of a specific gene measured under a certain condition. Some random variables, however, may specify unobserved quantities that are believed to influence the observable outcomes of a given experiment, such as which cellular processes were active at the time measurements were taken. The (directed or undirected) arcs of the graph specify the biological hypotheses about how observable and latent quantities influence one another. A set of constants underlying the distributions of the random variables completes the picture. These constants are referred to as {\em parameters} in the frequentist paradigm and as {\em hyper-parameters} in the Bayesian paradigm. (See \citet[pp.\ 185--189]{Wass:2004} for a discussion of when the distinction matters in practice, with examples.)

\begin{figure}[t!]
  \centering
   \includegraphics[width=0.75\textwidth]{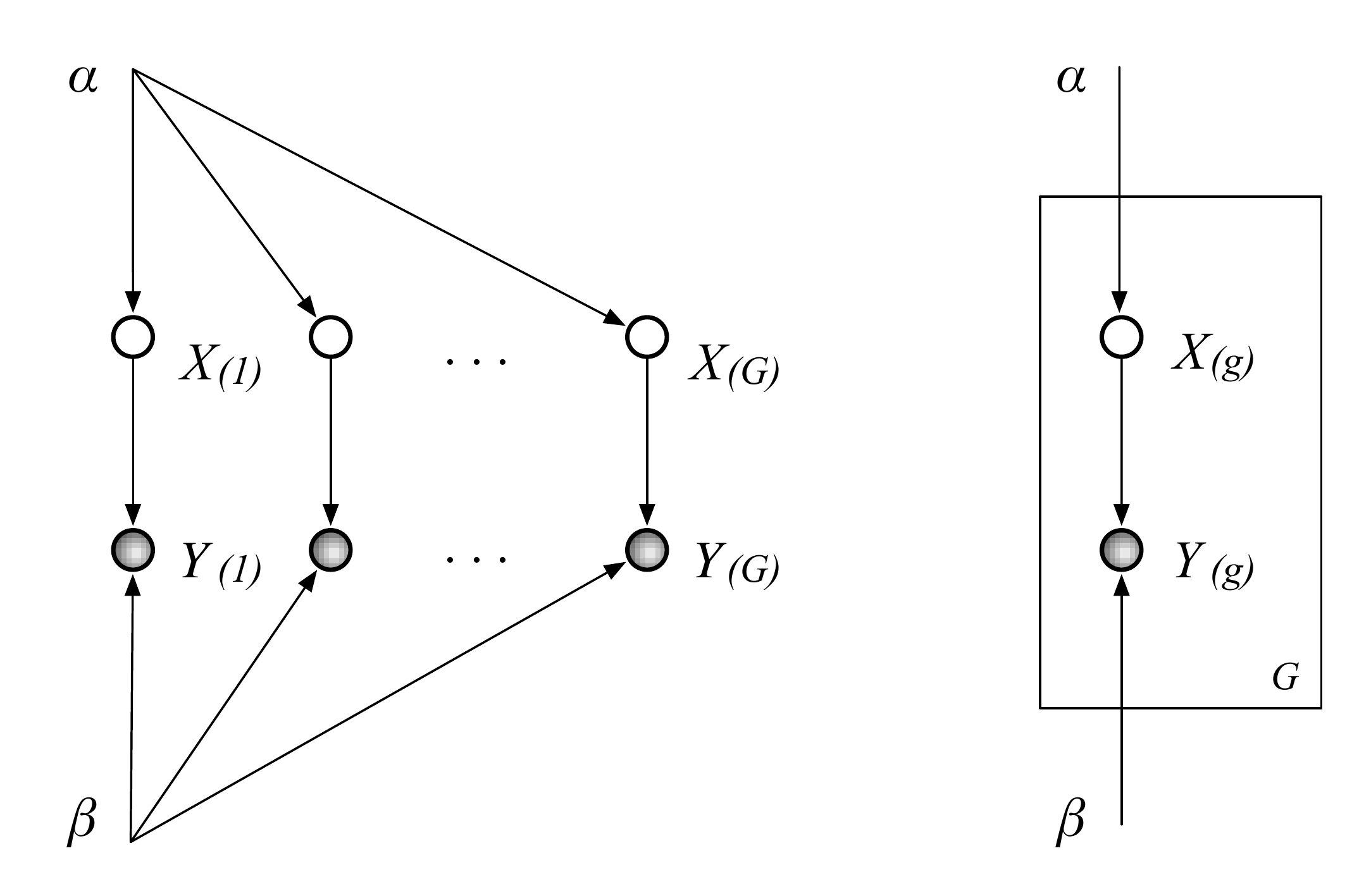}
 \caption{Two equivalent representations of the same probabilistic graphical model. The left panel shows the full model, and the right panel shows the same model expressed in compact form. Nodes denote random variables, observed random variables are shaded while latent random variables are not, edges denote possible dependences. The box in the right panel is called a {\em plate}; it denotes IID replicates.}
\label{fig:example}
\vskip 25pt
\end{figure}

 Figure \ref{fig:example} shows an example of a probabilistic graphical model for gene expression. (We note that there is a considerable overlap between the class of probabilistic graphical models and the class of Bayesian networks. A number of scholars choose to refer to PGMs that can be represented as directed acyclic graphs, with nodes corresponding to discrete-valued random variables, encoding observed measurements, and no latent variables as Bayesian networks.) The model encodes the intuition that the observed expression of a gene, $Y(g)$, depends on the latent functional process it is involved in, $X(g)$. The underlying constants, $(\alpha, \beta)$, control the probability that any given functional process is active and the probability of observing expression of a certain magnitude, respectively. The left panel shows the full model, and the right panel shows the same model expressed in compact form.

 The {\em likelihood function}, or the probability of the measurements given the underlying constants, is the main quantity of interest in PGMs. It summarizes how well the observations are explained by the specific PGM that is identified by a given value of the underlying constants. The likelihood can be computed using the structural hypotheses encoded by the graph, and the probability distributions specified for the nodes. Continuing the example, the likelihood corresponding to the model in Figure \ref{fig:example} is computed as follows,
\begin{eqnarray}
 \label{eq:lik_1}
  Pr \bigm({Y} \mid \alpha,\beta \bigm) 
   &=& \int_{\mathcal{X}} ~ Pr \bigm({Y}, {X} \mid \alpha,\beta \bigm) ~ d{X} \\
 \label{eq:lik_2}
   &=& \int_{\mathcal{X}} ~ \prod_{g=1}^G~\biggm[~Pr \bigm( Y(g) \mid X(g), \beta \bigm) \cdot ~ Pr \bigm( X(g) \mid \alpha \bigm)~\biggm] ~ d{X} \\
   &\triangleq&  \ell \bigm({Y} \mid \Theta \bigm),
\end{eqnarray}  
 for \( \Theta \triangleq (\alpha,\beta) \). The joint probability of measurements and latent variables given the underlying constants, that is, the integrand on the right-end side of Equation \ref{eq:lik_1}, is often referred to as the {\em complete likelihood function} in the literature---an important quantity in the statistical treatment of PGMs with latent variables.
  
\paragraph{Estimation and Inference.}

 A family of PGMs is {\em fit to the data} to find likely values for its underlying constants and likely distributions for its latent variables. This process boils down to an optimization problem where the objective function is based on the likelihood. Considered jointly, the estimation and inference tasks identify a specific model in the family of PGMs that is defined by the assumptions on the graph and the random variables, which successfully summarizes the variability of the observations. 

In the language of the statistical literature, we distinguish the task of {\em estimating} the underlying constants (i.e., the parameters in a frequentist statistical setting, or the hyper-parameters in a Bayesian statistical setting) of a probabilistic graphical model, from the task of {\em inferring} the distributions of the latent variables given the observations.
 Let us consider strategies to address the latter task first. The choice among the many strategies available is often informed by the complexity of the model, and in particular by whether the integral on the right-end side of Equation \ref{eq:lik_1} can be computed in closed form. Exact inference is available for models that belong to special families \citep{Jord:2004}. Focusing on the biology of the problem, however, often leads to a model structure and probabilistic specifications that cannot be subsumed under any special family. The likelihood is intractable, in many such cases---that is, the integral in Equation \ref{eq:lik_1} cannot be solved in closed form---and we resort to approximations. Below, we briefly survey the intuitions behind three popular strategies to perform approximate inference in PGMs: Monte Carlo Markov chains (sampling-based), and expectationÐmaximization (EM) and variational methods (optimization-based).
 
 Monte Carlo Markov chains (MCMC) techniques such as the Gibbs or Metropolis-Hastings samplers can be used to explore the joint posterior distribution of the latent variables \citep{Gelm:Carl:Ster:Rubi:1995,Robe;Case;2005}. 
 Although the likelihood is intractable, the complete likelihood $Pr \bigm({Y}, {X} \mid \alpha,\beta \bigm)$ can be easily computed for the large majority of PGMs. The main concept behind MCMC schemes is to work with the complete likelihood, and reduce the full joint posterior to lower-dimensional conditional distributionsÑon individual, or blocks of latent variablesÑthat we can sample from. Samples from the joint posterior are then obtained by composing conditional samples. The Gibbs sampler, for instance, requires that one can sample from all univariate, full-conditional distributions,
\begin{equation}
 Pr \bigm( X(g) \mid X_{(-g)}, Y, \alpha,\beta \bigm), \quad \hbox{for } ~ g=1,\dots,G,
\end{equation}
 where $X_{(-g)}$ is the collection of random variables $X$ without $X(g)$. The Metropolis-Hastings sampler requires that one can at least compute a quantity proportional to the desired posterior---samples are drawn from an arbitrary {\em proposal} distribution and are accepted or rejected using a formula that depends on the proposal. Other sampling-based algorithms such as particle filters can be used to perform inference in PGMs of sequential observations \citep{liu:2001}.

 The two alternatives to sampling we survey here aim at approximating the integral on right-end side of Equation \ref{eq:lik_1}. The main idea shared by both approaches is to find a lower bound for the likelihood, $\ell (Y \mid \Theta )$, making use of Jensen's inequality and of an arbitrary distribution on the latent variables $q(X)$,
\begin{eqnarray}
 \log ~ \ell \bigm({Y} \mid \Theta \bigm) & =    & \log \int_\mathcal{X} ~ Pr~(Y,X \mid\Theta) ~ dX \nonumber \\
                & =    & \log \int_\mathcal{X} ~ q(X) ~ \frac{Pr~(Y,X \mid\Theta)}{q(X)} ~ dX \qquad \hbox{(for any $q$)} \nonumber \\
                & \geq & \int_\mathcal{X} ~ q(X) ~ \log \frac{Pr~(Y,X \mid\Theta)}{q(X)} ~ dX \qquad \hbox{(Jensen's inequality)} \nonumber \\
                & =    & \mathbb{E}_q \bigm[ \log Pr~(Y,X \mid\Theta) - \log q(X) \bigm] ~ \triangleq ~ \mathcal{L}(q,\Theta)
\end{eqnarray}
 In EM, the lower bound $\mathcal{L}(q,\Theta)$ is then iteratively maximized with respect to $\Theta$, in the M step, and $q$ in the E step \citep{Demp:Lair:Rubi:1977}.  In particular, at the $t$-$th$ iteration of the E step the $q$ distribution must satisfy the following equation:
\begin{equation}
\label{eq:em_post}
 q^{(t)} = Pr \bigm(X \mid Y,\Theta^{(t-1)}\bigm).
\end{equation}
 That is, we set the arbitrary distribution $q$ equal to the posterior distribution of the latent variables given the data and the estimates of the  parameters at the previous iteration. Unfortunately, it is not always possible to express the distribution $q^{(t)}$ in Equation \ref{eq:em_post} in analytic form. 
 In such cases, a variational approximation to the EM algorithm \citep{Jord:Ghah:Jaak:Saul:1999} can be obtained by defining a parametric approximation to the posterior in Equation \ref{eq:em_post}, denoted by $\tilde q \triangleq q_\Delta (X)$, which involves an extra set of {\it variational parameters}, $\Delta$, and leads to an approximate lower bound for the likelihood $\mathcal{L}_\Delta (q,\Theta)$.
 At the $t$-$th$ iteration of the E step, we then minimize the Kullback-Leibler divergence between $q^{(t)}$ and $q^{(t)}_\Delta$, with respect to $\Delta$, using the data---this is equivalent to maximizing the approximate lower bound for the likelihood, $\mathcal{L}_\Delta (q,\Theta)$, with respect to $\Delta$.  The optimal parametric approximation can be thought of as an approximate posterior distribution for the latent variables in the sense that it depends on the data $Y$, although indirectly,
\[ 
 q^{(t)} \approx q_{\Delta^*(Y)}^{(t)} (X) = Pr~(X \mid Y).
\]
 
 Let us now return to the task of estimating the constants underlying a PGM; few established strategies exist. The estimates for the underlying constants may be chosen, for instance, to maximize the likelihood, or to match empirical and theoretical moments of the random variables that correspond to measurements \citep[pp.\ 120--124]{Wass:2004}. Alternatively, when the likelihood is too difficult or expensive to compute, an approximation, $\mathcal{L}_\Delta \approx \ell$, or a lower-bound, $\mathcal{L} \leq \ell$, for the likelihood can be used as a surrogate. These alternatives and others are sometimes referred to as empirical Bayes estimates in the context of non-trivial probabilistic graphical models \citep[Chapter 3]{Carl:Loui:2005}.

 Popular software packages that implement a language to specify and fit PGMs are available. For MCMC see BUGS \citep{Lunn:Thom:Best:Spie:2000}; for variational inference see VIBES \citep{Bish;Spie;Winn;2003}.
 
\paragraph{Applications.}

 With the technical machinery we just introduced, we are now ready to bring the biological intuition back into the picture.
 Let us continue with the transcriptional regulation example. In the PGM of Figure \ref{fig:example}, the expression of gene $g$ may be encoded by a real-valued random variable $Y(g)$. The mixed membership of gene $g$ to non-observable biological contexts may be encoded by the nonzero components of a latent random vector, $X(g)$. The number of latent biological contexts we ask the PGM to infer, denoted by $K$, is an important quantity in this model, which we discuss later---briefly, $K$ specifies the {\em dimensionality} of this PGM, that is, the number of components of the vector-valued latent variables, $X(g)$.  The two constants $(\alpha,\beta)$ may be used to encode biological constraints. For instance, $\alpha$ may be used to introduce a notion of biological parsimony in the form of a probabilistic (soft) constraint on the number of biological contexts each gene may participate in, and $\beta$ may be used to specify gene expression patterns in the form of differential expression levels across those experimental conditions for which microarray measurements were taken---alternative pattern specifications and parameterizations exist \citep{Airo:Fien:Xing:2006}.
 For any given number of latent biological contexts, $K$, the PGM is fit to the data. Estimation and inference will assign numerical values to the unknown quantities $(\vec X,\alpha,\beta)$.
 These quantities provide us with {\em model-based} and {\em observation-induced} summaries of the data. In the example, for instance, while $\beta$ summarizes gene expression patterns that summarize the main trends of transcription in a collection of microarrays, the values assigned to the latent variables, $X(g)$, provide gene-specific information that can be used for making fine-grained predictions.

 In the last last stage of the analysis, we assess of the biological relevance of the patterns we inferred from the data (such as the biological contexts, or gene co-expression patterns, in the example) to make sure the model is capturing the signal we set out to capture, and we use the inferred patterns to gain insights into the problem.
 Assessment of biological relevance can be qualitative or quantitative.
 Qualitative methods such as visual inspection are typically useful for focused scientific endeavors; for instance, whenever biological problem targets a small set of genes or a specific cellular process or component, or a signaling pathway.
 Quantitative methods are necessary for genome-wide scientific endeavors, and typically rely on knoledge based repositories and ontologies \citep[such as gene ontology,][]{Ashb:etal:2000} and bioinformatics tools to carry out the evaluation \citep[e.g.,][]{Boyl:Weng:Goll:Jin:2004,Myer:Barr:Hibb:Hutt:2006}.
 Arguably, in any given application, the more interpretable the patterns are, in terms of functional processes and other biological concepts of interest, the better the family of PGMs captures some aspects of biology that may be relevant for the understanding of the phenomenon under investigation, and that are not directly measurable with experimental techniques.

 Moving a step forward, the goodness of model fit is often taken as a measure of {\em how well} the data support structural biological hypotheses encoded by the cartoon model of biology that was used to posit a given family of PGMs. 
 Measures of goodness of model fit include the Bayesian information criterion, the held-out likelihood obtained using bootstrap or cross-validation techniques, measures of predictive power such as the predictive $R^2$ in linear regression, or other quantities, depending on the goals of the analysis. 
 (These measures can also be used to select the dimensionality, $K$, of the PGM in the example.)
 The goodness of fit, along with the substantive value of the inferred patterns, should inform a critical review of the biological assumptions underlying the initial cartoon model, and possibly suggest new hypotheses---testable either with new statistical analyses, or with new experimental probes at the bench. In this sense, probabilistic graphical models contribute to an iterative process of scientific discovery, where statistical and biological thinking are intertwined as both cause and effect.

 There is a rich history of applied research that leverages the probabilistic graphical models approach outlined above to problems in the biological sciences.
 It includes a model for inferring the ancestral population structure of individuals starting from a collection of multilocus genotype measurements  \citep{Prit:Step:Donn:2000,Sohn:Xing:2007} and a model for inferring HIV mutation patterns from longitudinal clonal sequence data  \citep{Beer:Drto:2007}; the former model is closely related to the classic probabilistic graphical models to infer phylogenetic trees  \citep{Fels:1981,Fels:Chur:1996} and to recent extensions, in particular, that take into account the dependence among the bases at neighboring sites  \citep{McAu:Pach:Jord:2004,Siep:Haus:2004}.
 Models for sequence analysis are well established in the community  \citep{Durb:Eddy:Krog:Mitc:1998,Xing:Karp:2004}; more recently, the connection between sequence information and gene expression has been investigated using probabilistic graphical models as well  \citep{Sega:Yele:Koll:2003,Beer:Tava:2004}.
 Other applications of this research include: a model for predicting the clinical status of breast cancer using gene expression profiles  \citep{West:Blan:Dres:Huan:2001}; a model for facilitating content browsing of biomedical literature about the nematode {\em Caenorhabditis elegans} \citep{Blei:Fran:Jord:Mian:2006};a model for inferring the location of chromosome aberrations from array-based comparative genomic hybridization measurements  \citep{Myer:Dunh:Kung:Troy:2004}; and an extension that leverages array-based comparative genomic hybridization profiles from multiple individuals to recover shared aberration patterns \citep{Shah:Lam:Ng:Murp:2007}; a model for reconstructing features of the internal organization of the cell from the nested structure of observed perturbation effects, such as those measured via high-dimensional phenotype screens  \citep{Mark:Kost:Troy:Span:2007}; a model for inferring proteins' multiple functional roles from a large collection of manually curated protein interactions, as well as cross-talk patterns among proteins that participate in distinct functional processes \citep{Airo:Blei:Fien:Xing:2006a}; and a model for inferring temporal patterns of coexpressed genes from time-course expression data measured via SAGE and microarray technologies \citep{Airo:Fien:Xing:2006}.

 Note that the graphical representation of a family of PGMs goes only so far in specifying the model; itÕs informative, but not exhaustive. Probabilistic assumptions and some features of the sampling scheme cannot be specified by the graph. Such subtle variants typically make a significant difference in applications. 
 
\paragraph{Conclusions.}

 Probabilistic graphical models offer a common conceptual architecture where biological and mathematical objects can be expressed with a common, intuitive formalism. This enables effective communication between scientists across the mathematical divide by fostering substantive debate in the context of a scientific problem, and ultimately facilitates the joint development of statistical and computational tools for quantitative data analysis.
 In other words, probabilistic graphical models provide a bridge between biology and statistical computations. These models recently earned a spot at the center stage of modern (computational) biology by furthering our ability to probe data for biological hypotheses, and will undoubtedly play an important role in resolving many intriguing conundrums in the biological sciences, in the future.

\paragraph{Acknowledgments.}

 This research was partly supported by United States National Institute of General Medical Sciences Center of Excellence grant P50 GM071508, by National Science Foundation grants DBI-0546275 and IIS-0513552, and by National Institutes of Health grant R01 GM071966.
 The author thanks Florian Markowetz, Chad Myers, David Hess, and Olga Troyanskaya at Princeton, and Eric Xing at Carnegie Mellon, for comments on an early draft of this manuscript.

\bibliographystyle{plainnat}
\renewcommand\refname{\normalsize Essential Bibliography.}

\end{document}